\documentclass{ws-ijmpcs}

\begin{document}

\markboth{Hai-Yang Cheng}
{Theoretical Review on Hadronic $D/D_s$ Decays}

%
\catchline{}{}{}{}{}
%

\title{Theoretical Review on Hadronic $D/D_s$ Decays
}

\author{Hai-Yang Cheng}

\address{Institute of Physics, Academia Sinica,
Taipei, Taiwan 115, R.O.C.
\\
phcheng@phys.sinica.edu.tw}

\maketitle


\begin{abstract}
A brief review on the theoretical status of hadronic charm decays is presented. We emphasize the analyses based on the diagrammatic approach and the connection of weak annihilation with final-state rescattering. The topics on the $D^0$-$\overline D^0$ mixing and the baryonic charm decay $D_s^+\to p\bar n$ are sketched.

\keywords{D meson; hadronic decays; mixing.}
\end{abstract}

\ccode{PACS numbers: 13.25.Ft, 11.30.Hv, 12.15.Ff}

\section{Introduction}	
In $B$ physics, there exist several QCD-inspired approaches describing the nonleptonic $B$ decays, such as QCD factorization (QCDF), pQCD and soft collinear effective theory. However, this is not the case in the $D$ sector. Although charm has been with us for 35 years, until today a theoretical description of the underlying mechanism for exclusive hadronic $D$ decays based on QCD is still not yet available.  This has to do with the mass of the charm quark, of order 1.5 GeV.  It is not heavy enough to allow for a sensible heavy quark expansion and light enough for the application of chiral perturbation theory. After more than three decades, we are still back to square one: Besides the effective short-distance Hamiltonian we need to rely on naive factorization, or the improved version of factorization such as the generalized factorization known as the Bauer-Stech-Wirbel model\cite{BSW} or the factorization based on $1/N_c$ expansion. In the past 15 years or so, people have tried to apply pQCD or QCDF to hadronic charm decays. However, it does not make much sense to generalize these approaches to charm decays as the $1/m_c$ power corrections are so large that the heavy quark expansion in $1/m_c$ is beyond control.

There is another powerful tool which provides a model-independent analysis of the charm decays, namely, the diagrammatic approach. It is complementary to the factorization approach. Analysis based on the flavor-diagram approach indicates a sizable weak annihilation ($W$-exchange or $W$-annihilation) topological amplitude with a large phase relative to the tree amplitude. Since weak annihilation and final-state interactions (FSIs) are both of order $1/m_c$ in the heavy quark limit, this means FSIs could play an essential role in charm decays. Indeed, we shall see that weak annihilation contributions arise mainly from final-state rescattering. This explains why an approach based on heavy quark expansion in $1/m_c$ is not suitable for charm decays.

\section{Diagrammatic approach \label{sec:topology}}
The
two-body nonleptonic weak decays of heavy mesons can be analyzed in terms of six distinct quark diagrams\cite{Chau,CC86}: $T$, the color-allowed external $W$-emission tree diagram; $C$, the color-suppressed internal $W$-emission diagram; $E$, the $W$-exchange diagram; $A$, the $W$-annihilation diagram; $P$, the horizontal $W$-loop diagram; and $V$, the vertical $W$-loop diagram.
(The one-gluon exchange approximation of the $P$ graph is the so-called ``penguin diagram''.)  It should be stressed that these diagrams are classified according to the topologies of weak interactions with all strong interaction effects encoded, and hence they are {\it not} Feynman graphs.    Therefore, analyses of topological graphs can provide information on final-state interactions.

\subsection{$D\to PP$ decays}

The first study of topological amplitudes in charm decays was due to Rosner\cite{Rosner99}.
Using the most recent CLEO measurements\cite{CLEOPP09} for $D\to PP$ decays, the reduced quark-graph amplitudes $T,C,E,A$ are extracted from the Cabibbo-favored (CF) $D\to PP$ decays to be (in units of $10^{-6}$ GeV)\cite{CC:charm,RosnerPP09}
\begin{eqnarray} \label{eq:PP1}
&& T=3.14\pm0.06, \qquad\qquad\qquad\quad
C=(2.61\pm0.08)\,e^{-i(152\pm1)^\circ}, \nonumber \\
&&  E=(1.53^{+0.07}_{-0.08})\,e^{i(122\pm2)^\circ},
\qquad\quad  A=(0.39^{+0.13}_{-0.09})\,e^{i(31^{+20}_{-33})^\circ}
\end{eqnarray}
for $\phi=40.4^\circ$, the $\eta-\eta'$ mixing angle defined in terms of the flavor states $\eta_q=(u\bar u+d\bar d)/\sqrt{2}$ and $\eta_s=s\bar s$. The fitted $\chi^2$ value is 0.29 per degree of freedom with quality $59.2\%$.

From Eq. (\ref{eq:PP1}) we see that the color-suppressed amplitude $C$ not only is comparable to the tree amplitude $T$ in magnitude but also has a large strong phase relative to $T$. (It is $180^\circ$ in naive factorization.) The $W$-exchange $E$ is sizable with a large phase of order $120^\circ$. Since $W$-exchange is of order $1/m_c$ in the heavy quark limit, this means that $1/m_c$ corrections are very important in charm decays.  Finally, we see that $W$-annihilation is substantially smaller than $W$-exchange and almost perpendicular to $E$.

\begin{figure}[t]
\begin{center}
\includegraphics[width=0.70\textwidth]{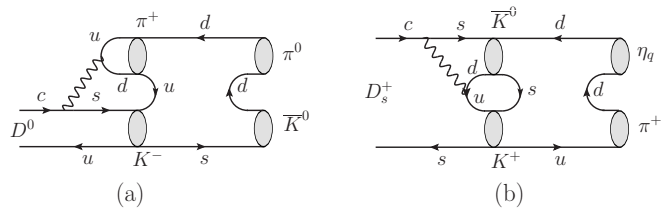}
\vspace{0.0cm}
\caption{Long-distance contributions to $D^0\to \overline K^0\pi^0$
    and $D_s^+\to \pi^+\eta_q$ from
    the color-allowed $D^0\to K^-\pi^+$ and color-suppressed $D_s^+\to K^+\overline{K}^0$ decays, respectively, followed by a resonant-like rescattering. While (a)
    has the same topology as the $W$-exchange graph, (b) mimics
    the $W$-annihilation amplitude.} \label{fig:DFSI} \end{center}
\end{figure}

In naive factorization, the factorizable weak annihilation amplitudes are usually assumed to be negligible as they are helicity suppressed or, equivalently, the form factors are suppressed at large $q^2=m_D^2$. At first glance, it appears that the factorizable weak annihilation amplitudes are too small to be consistent with experiment at all.  However, in the diagrammatic approach here, the topological amplitudes $E$ and $A$ do receive contributions from the tree and color-suppressed amplitudes $T$ and $C$, respectively, via final-state rescattering, as illustrated in Fig.~\ref{fig:DFSI}. Therefore, even if the short-distance weak annihilation vanishes, a long-distance weak annihilation can be induced via inelastic FSIs.

\begin{figure}[t]
\begin{center}
\includegraphics[width=0.65\textwidth]{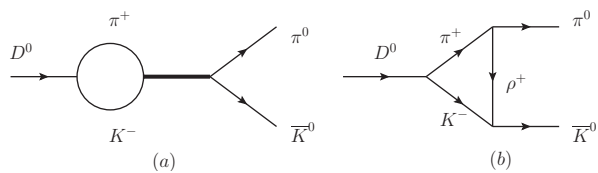}
\vspace{0.0cm}
\caption{Manifestation of Fig. \ref{fig:DFSI}(a) as the
    long-distance $s$- and $t$-channel contributions to the
    $W$-exchange amplitude in $D^0\to \overline K^0\pi^0$. The thick
    line in (a) represents a resonance.} \label{fig:Wexchange} \end{center}
\end{figure}

Since FSIs are nonperturbative in nature, in principle it is
extremely difficult to calculate their effects. It is customary to
consider the final-state rescattering at the hadron level. For example, Fig. \ref{fig:DFSI}(a) is manifested
at the hadron level as Fig. \ref{fig:Wexchange}: an
$s$-channel scalar particle exchange with the quark content
$(s\bar d)$ and a $t$-channel contribution with a $\rho$
particle exchange. The next question is which contribution dominates. One thing unique to charm decays is that an
abundant spectrum of resonances is known to exist at energies
close to the mass of the charmed meson. As
emphasized in \cite{Zen,Weinberg}, most of the properties of
resonances follow from unitarity alone, without regard to the
dynamical mechanism that produces the resonance.
\footnote{Final-state interaction effects have been studied by Buccella et al.\cite{Buccella} by assuming that FSIs are dominated by nearby resonances, similar to the work of \cite{Zen,Chenga1a2}. The major uncertainties there arise from the unknown masses and widths of the resonances at energies near the charmed meson mass. Moreover, as stressed in \cite{Chenga1a2} (see also Sec. 2.2 below), weak annihilation in $VP$ systems receives little contributions from resonant FSIs.}
As
shown in \cite{Zen,Chenga1a2}, the effect of resonance-induced
FSIs [Fig. 2(a)] can be described in a model-independent manner in
terms of the mass and width of the nearby resonances. It is found that weak annihilation amplitudes are modified by resonant FSIs as\cite{Zen,Chenga1a2}
\begin{eqnarray} \label{E}
 E=e+(e^{2i\delta_r}-1)\left(e+{T\over 3}\right), \qquad
 A=a+(a^{2i\delta_{r'}}-1)\left(a+{C\over 3}\right),
\end{eqnarray}
with $e^{2i\delta_{r^{(')}}}=1-i\,{\Gamma_{R^{(')}}\over m_D-m_{R^{(')}}+i\Gamma_{R^{(')}}/2}$, where  the $W$-exchange amplitude $E$ and $W$-annihilation $A$ before resonant FSIs are denoted by $e$ and $a$, respectively. Therefore, even if the short-distance weak annihilation is turned off, a long-distance $W$-exchange ($W$-annihilation) contribution still can be induced from the tree amplitude $T$ ($C$) via FSI rescattering in resonance formation. To see the importance of resonant FSIs, consider the scalar resonance
$K^*_0(1950)$ with a mass $1945\pm
10\pm 20$ MeV and a width $201\pm 34\pm 79$ MeV, contributing to the $W$-exchange in $D\to K\pi,K\eta$.
Assuming
$e=0$ in Eq. (\ref{E}), we obtain
$E=1.68\times 10^{-6}\,{\rm exp}(i143^\circ)\,{\rm GeV}$,
which is close to the ``experimental" value of $E$ given in Eq. (\ref{eq:PP1}). This implies that {\it weak annihilation topologies in $D\to PP$ decays are dominated by nearby  resonances via final-state rescattering.}

Under the flavor SU(3) symmetry, one can use the topological amplitudes extracted from CF modes to predict the rates for singly Cabibbo-suppressed (SCS) and doubly Cabibbo-suppressed (DCS) decays. In general, the agreement with experiment is good except some discrepancies in SCS decays. For example, the predicted rates for $\pi^+\pi^-$ and $\pi^0\pi^0$ are too large, while those for $K^+K^-$, $\pi^+\pi^0$, $\pi^+\eta$, $\pi^+\eta'$, $K^+\eta$ and $K^+\eta'$ are too small compared to experiment\cite{CC:charm}. We find that part of the SU(3) breaking effects can be accounted for by SU(3) symmetry violation manifested in the color-allowed and color-suppressed tree amplitudes.  However, in other cases such as the ratio $R=\Gamma(D^0\to K^+K^-)/\Gamma(D^0\to\pi^+\pi^-)$, SU(3) breaking in spectator amplitudes leads to $R=1.6$ and
is still not sufficient to explain the observed value of $R\approx 2.8$.  This calls for the consideration of SU(3) violation in the $W$-exchange amplitudes. We argue that the long-distance resonant contribution through the nearby state $f_0(1710)$ can naturally explain why $D^0$ decays more copiously to $K^+ K^-$ than $\pi^+ \pi^-$ through the $W$-exchange topology\cite{CC:charm}. This has to do with the dominance of the scalar glueball content of $f_0(1710)$ and the chiral-suppression effect in the decay of a scalar glueball into two pseudoscalar mesons. The same FSI through the $f_0(1710)$ pole contribution also explains the occurrence of $D^0\to K^0\bar K^0$.  In short, the long-standing puzzle for $D^0\to K^+K^-/\pi^+\pi^-$ is now understood.

\subsection{$D\to VP$ decays}
Since the spectator quark of the charmed meson may end up in the pseudoscalar or vector meson, there exist two different types of reduced amplitudes $T_P$ and $T_V$ for the spectator topology and likewise for weak annihilation. A fit to the data of CF $D\to VP$ decays gives the best solutions\cite{CC:charm,RosnerVP} (in units of $10^{-6}$)
\begin{eqnarray}
&& T_V=4.16^{+0.16}_{-0.17},
\quad C_P=(5.14^{+0.30}_{-0.33})\,e^{-i(162\pm3)^\circ},
\quad E_P=(3.09\pm0.11)e^{-i(193\pm5)^\circ} , \nonumber \\
&& T_P=8.11^{+0.32}_{-0.43},
\quad C_V=(4.15^{+0.34}_{-0.57})\,e^{i(164^{+36}_{-10})^\circ},
\quad E_V=(1.51^{+0.97}_{-0.69})e^{-i(124^{+57}_{-26})^\circ} . \end{eqnarray}
We didn't find solutions for $A_P$ and $A_V$. The topological amplitude expressions of $D_s^+\to \pi^+\rho^0$ and $D_s^+\to\pi^+\omega$ in units of $V_{cs}^*V_{ud}$ are given by
\begin{eqnarray}
{\cal A}(D_s^+\to \pi^+\rho^0) = {1\over\sqrt{2}}(A_V-A_P), \quad
{\cal A}(D_s^+\to \pi^+\omega) = {1\over\sqrt{2}}(A_V+A_P).
\end{eqnarray}
Under the $G$-parity argument, the decay $D_s^+\to \pi^+\omega$ is prohibited via direct or resonance-induced $W$-annihilation; that is, $A_P=-A_V$.
Indeed, a general consideration of resonant FSIs gives the relations\cite{Zen,Chenga1a2}
\begin{eqnarray}
 A_P^r+A_V^r = a_P+a_V,~~
 A_P^r-A_V^r = a_P-a_V+(e^{2i\delta_r}-1)\left(a_P-a_V+{1\over
 3}(C_P-C_V)\right).
\end{eqnarray}
The above relation shows that $A_P+A_V$ does not receive any $\bar qq'$ resonance ({\it e.g.}, the $0^-$ resonance $\pi(1800)$) contributions.   Experimentally, however, it is the other way around\cite{PDG}: ${\cal B}(D_s^+\to\pi^+\omega) = (2.3\pm 0.6)\times 10^{-3}$ and ${\cal B}(D_s^+\to\pi^+\rho^0)=(2.0\pm1.2)\times 10^{-4}$. To resolve this puzzle, we notice that there are long-distance final-state rescattering contributions to $D_s^+\to \pi^+ \omega$ allowed by $G$-parity conservation.  A nice example is the contribution from the weak decay $D_s^+\to\rho^+\eta^{(')}$ followed by quark exchange. The rescattering of $\rho^+\eta^{(')}$ into $\pi^+\rho^0$ is prohibited by the $G$-parity selection rule. Consequently, $A_P+A_V=A_P^e+A_V^e=2A_P^r$, where the superscript $e$ indicates final-state rescattering via quark exchange, and $A_P-A_V=A_P^r-A_V^r$. Since $D_s^+\to\rho^+\eta$ has the largest rate among the CF $D_s^+\to VP$ decays,  it is conceivable that $D_s^+\to \pi^+\omega$ can be produced via FSIs at the $10^{-3}$ level as its branching fraction. Therefore, {\it contrary to the $PP$ sector, weak annihilation in $VP$ systems is dominated by final-state rescattering via quark exchange.}

In principle, the annihilation amplitudes can be determined from the four decays $\bar K^{*0}K^+$, $\bar K^0K^{*+}$, $\pi^+\omega$ and $\pi^+\rho^0$. However since $|C_P|>|C_V|\gg |A_P|, |A_V|$, it is not possible to find a nice fit to the above-mentioned four decays  simultaneously. This leads to a contradiction with data, as $\Gamma(D_s^+ \to {\bar K}^0 K^{*+}) > \Gamma(D_s^+ \to {\bar K}^{*0} K^+)$ while the former is dominated by $C_V$ and the latter by $C_P$.  So a full determination of the $A_P$ and $A_V$ amplitudes still await more precise data on the related modes. We conjecture that the currently quoted experimental results for both $D_s^+\to\bar K^0K^{*+}$ and $D_s^+\to \rho^+\eta'$ are overestimated and problematic.

\section{Applications}
There are two interesting applications of the diagrammatic approach:
\vskip 0.1 cm
\underline{$D^0$-$\overline D^0$ mixing}~~~
The $D^0$-$\overline D^0$ mixing is governed by the mixing parameters $x=(m_1-m_2)/\Gamma$ and $y=(\Gamma_1-\Gamma_2)/(2\Gamma)$ for the mass eigenstates $D_1$ and $D_2$. It is known that the short-distance contribution to the mixing parameters is very small\cite{Cheng:1982}, of order $10^{-6}$. On the theoretical side, there are two scenarios: the inclusive approach relied on $1/m_c$ expansion (see e.g. \cite{Lenz} for a recent study) and the exclusive approach with all intermediate states summed over. In the well-known paper \cite{Falk:y}, only the SU(3) breaking effect in phase space was considered for the estimate of $y$. Consequently, the previous estimate of mixing parameters is subject to large uncertainties. We believe that a better approach is to concentrate on 2-body decays and rely more on the data and less on theory. This is because the measured 2-body decays account for about 75\% of hadronic rates of $D$ mesons. For $PP$ and $VP$ modes, data with good precision for CF and SCS decays are now available. For as-yet unmeasured DCS modes, their rates can be determined from the diagrammatic approach. We obtain\cite{CC:mixing} $x_{PP+VP}=(0.10\pm0.02)\%$ and $y_{PP+VP}=(0.36\pm0.26)\%$. Since $PP$ and $VP$ final states account for nearly half of the hadronic width of $D^0$, it is conceivable that when all hadronic states are summed over, one could have $x\sim (0.2-0.4)\%$ and $y\sim (0.5-0.7)\%$. They are consistent with the recent BaBar measurement\cite{BaBar:DD2010} $x=(1.6\pm2.3\pm1.2\pm0.8)\times 10^{-3}$ and
$y=(5.7\pm2.0\pm1.3\pm0.7)\times 10^{-3}$. We believe that our estimate of the mixing parameters is more realistic and reliable.

\vskip 0.1 cm
\underline{Baryonic decay: $D_s^+\to p\bar n$}~~~
$D_s^+\to p\bar n$ is the only allowed baryonic charm decay and it proceeds via $W$-annihilation. Since the factorizable decay amplitude vanishes in the chiral limit, its branching fraction is very small, of order $10^{-6}$. This mode was first observed by CLEO\cite{CLEO} with the result ${\cal B}(D_s^+\to p\bar n)=(1.30\pm0.36^{+0.12}_{-0.16})\times 10^{-3}$. It receives long-distance contributions through final-state scattering of tree and color-suppressed amplitudes. Assuming that the long-distance enhancement of $W$-annihilation in the baryonic $D_s^+$ decay is similar to that in the meson sector, where the latter can be obtained from the analysis of the diagrammatic approach, we find that $D_s^+\to p\bar n$ becomes visible\cite{Chen:2008pf}. The observation of this baryonic charm decay implies the dynamical enhancement of $W$-annihilation in the $D_s^+$ decay.

\section*{Acknowledgments}

The author is grateful to Cheng-Wei Chiang for fruitful collaboration and to Hai-Bo Li for organizing this wonderful workshop.



\begin{thebibliography}{0}    


\bibitem{BSW}  M. Bauer, B. Stech, and M. Wirbel, {\it Z. Phys. C} {\bf 34}, 103 (1987).

\bibitem{Chau} L.-L. Chau Wang, p. 419-431 in AIP Conference Proceedings 72 (1980), Weak Interactions as Probes of Unification (edited by G.B. Collins, L.N.
Chang and J.R. Ficenec), and p.1218-1232 in Proceedings of the 1980 Guangzhou Conference on Theoretical Particle Physics (Science Press, Beijing, China, 1980, distributed by Van Nostrand Reinhold company); L.-L.
Chau, {\it Phys. Rep.} 95, 1 (1983).


\bibitem{CC86} L. L. Chau and H.Y. Cheng, Phys. Rev. Lett. {\bf 56}, 1655 (1986); {\it Phys. Rev. D} {\bf 36}, 137 (1987); {\it Phys. Lett. B}
{\bf 222}, 285 (1989).

\bibitem{Rosner99} J.L. Rosner, {\it Phys. Rev. D} {\bf 60}, 114026 (1999).

\bibitem{CLEOPP09}
  H.~Mendez {\it et al.}  [CLEO Collaboration],
 {\it Phys.\ Rev.\  D} {\bf 81}, 052013 (2010)

\bibitem{CC:charm}
  H.~Y.~Cheng and C.~W.~Chiang,
  {\it Phys. Rev. D} {\bf 81}, 074021 (2010).

\bibitem{RosnerPP09}
  B.~Bhattacharya and J.~L.~Rosner,
 {\it Phys.\ Rev.\  D} {\bf 81}, 014026 (2010).

\bibitem{Zen} P. \.Zenczykowski, {\it Acta Phys. Polon. B} {\bf 28}, 1605 (1997).

\bibitem{Weinberg} S. Weinberg, {\it The Quantum Theory of Fields,
Volume I} (Cambridge, 1995), Sec. 3.8.

\bibitem{Buccella} F. Buccella, M. Lusignoli, G. Miele, A.
Pugliese, and P. Santorelli, {\it Phys. Rev. D} {\bf 51}, 3478 (1995); F. Buccella, M. Lusignoli, and A. Pugliese, {\it Phys. Lett. B} {\bf
379}, 249 (1996).

\bibitem{Chenga1a2} H. Y. Cheng, {\it Eur. Phys. J. C} {\bf 26}, 551 (2003).

\bibitem{RosnerVP}
  B.~Bhattacharya and J.~L.~Rosner,
  {\it Phys.\ Rev.\  D} {\bf 79}, 034016 (2009).

\bibitem{PDG}
K. Nakamura {\it et al.} (Particle Data Group), {\it J. Phys. G} {\bf 37}, 075021 (2010).

\bibitem{Cheng:1982}
  H.~Y.~Cheng,
  {\it Phys.\ Rev.\  D} {\bf 26}, 143 (1982);
  A.~Datta and D.~Kumbhakar,
  {\it Z.\ Phys.\  C} {\bf 27}, 515 (1985).

\bibitem{Lenz}
  M.~Bobrowski, A.~Lenz, J.~Riedl and J.~Rohrwild,
  {\it JHEP} {\bf 1003}, 009 (2010)

\bibitem{Falk:y}
  A.~F.~Falk, Y.~Grossman, Z.~Ligeti and A.~A.~Petrov,
  {\it Phys.\ Rev.\  D} {\bf 65}, 054034 (2002).

\bibitem{CC:mixing}
  H.~Y.~Cheng and C.~W.~Chiang,
  {\it Phys.\ Rev.\  D} {\bf 81}, 114020 (2010).

\bibitem{BaBar:DD2010}
  P. del Amo Sanchez {\it et al.}  [BABAR Collaboration],
 {\it  Phys.\ Rev.\ Lett.\ } {\bf 105}, 081803 (2010).

\bibitem{CLEO}
  S.~B.~Athar {\it et al.}  [CLEO Collaboration],
  {\it Phys.\ Rev.\ Lett.\ } {\bf 100}, 181802 (2008).

\bibitem{Chen:2008pf}
  C.~H.~Chen, H.~Y.~Cheng and Y.~K.~Hsiao,
  {\it Phys.\ Lett.\  B} {\bf 663}, 326 (2008).

\end{thebibliography}
\end{document}